\documentclass[a4paper,fleqn,usenatbib]{mnras}

\usepackage{newtxtext,newtxmath}

\usepackage[T1]{fontenc}
\usepackage{ae,aecompl}

\usepackage{graphicx}	
\usepackage{amsmath}	
\usepackage{amssymb}	
\usepackage{pdflscape}

\title[Partial banana mapping]{The partial banana mapping: a robust linear method for impact probability estimation}

\author[D. E. Vavilov]{
	Dmitrii E. Vavilov\thanks{E-mail: vavilov@iaaras.ru}
	\\
	Institute of Applied Astronomy of the Russian Academy of Sciences, Kutuzova emb., St. Petersburg 191187, Russian Federation
}

\date{Accepted XXX. Received YYY; in original form ZZZ}

\pubyear{2019}

\begin{document}
\label{firstpage}
\pagerange{\pageref{firstpage}--\pageref{lastpage}}
\maketitle

\begin{abstract}

This paper presents a robust linear method for impact probability estimation of near-Earth asteroids with the Earth. This method is a significantly modified and improved method, which uses a special curvilinear coordinate system associated with the nominal orbit of an asteroid. One of the coordinates of this system is the mean anomaly in the osculating orbit of an asteroid. A normal distribution of errors of coordinates and velocities of this system is assumed. Because of the usage of the curvilinear coordinate system,  the fact that the confidence region is curved and stretched mainly along the nominal asteroid orbit is taken into account. On the main axis of the curvilinear confidence ellipsoid the virtual asteroid, which is the closest to the Earth, is found. The part of the curvilinear confidence ellipsoid, around the found virtual asteroid, is obtained and mapped on to its target plane. The impact probability is calculated as the probability of the asteroid being in the region of the found virtual asteroid multiplied by the probability of a collision of the found virtual asteroid with the Earth. This approach is shown to give more accurate and trustworthy results than the target plane method.

\end{abstract}




\section{Introduction}

Since the orbit of an asteroid is not known precisely there is a set of possible orbits (virtual asteroids~\citep{2000Icar..145...12M}), which are in general close to each other at the epoch of observations but will be significantly separated in the future, especially in Cartesian coordinates. All of these orbits are more or less likely depending on the value of the probability density function. 
Some of these virtual asteroids (VAs) can collide with the Earth thereby producing an impact probability of the considered asteroid.

In general, at the epoch of observations the probability density function of orbital parameters is assumed to be normal.  Modelling the evolution of this function with time is a difficult conundrum but is required to estimate the impact probability. In linear methods for impact probability computation the errors of orbital parameters are assumed to be linearly related to the ones at the epoch of observations making them have a normal distribution at all considered times.

Paul \citet{1993BAAS...25.1236C} was the first who introduced linear methods for computing impact probabilities of asteroids with the Earth and suggested using a target plane. The method was immediately tested for prediction a collision of comet Shoemaker-Levy~9 with Jupiter estimating the probability as 64\% \citep{1993IAUC.5807....2C}. However, this method has several limitations. First of all, a linear relation between orbital parameters' errors can be broken by, for instance, a close approach with a massive body leading the method to fail. This is a drawback of all linear methods. Secondly, the target plane method is used when  {the asteroid on its nominal orbit (orbit which was derived through orbital fitting by least-squares method} \citep{1809tmcc.book.....G})
comes close to the Earth and the uncertainty region is assumed to be an ellipsoid. While actually this region is mostly stretched along the nominal orbit of the asteroid due to distinctions in the mean motions of VAs, making the region appear like a curved ellipsoid or 'bananoid'.  This fact makes a  collision possible even if the nominal asteroid is far from the Earth.

In order to take this fact into account \citet{2015MNRAS.446..705V} introduced a new curvilinear coordinate system related to the nominal asteroid's orbit. The actual distribution of VAs can be well approximated by a normal law of errors of coordinates and velocities in this system and the 'bananoid' is defined by the equation of ellipsoid in this coordinate system. An impact probability is calculated as a probability that at time $t$ the asteroid is located closer to the Earth's centre than the Earth's radius by computing a three-dimensional integral of normal law. Even though this approach was showed to be more robust than the target plane method it has its own disadvantages. The probability is computed for a certain time hence we must know the collision time with an accuracy of about 1 hour. In practice, a set of probabilities with small time step is computed in a several days vicinity of the epoch when the Earth is closest to the asteroid's orbit. Also the so computed probability is not a cumulative probability around some date which is a more interesting value. Besides, calculating a three-dimensional integral for a set of possible collisions is time consuming. The above mentioned drawbacks became a motive for improving the method that is done in the current work.

Even though there are non-linear methods for impact probability computation such as the Monte-Carlo method, Line of Variation sampling method (LOV) \citep{2002aste.book...55M,2005A&A...431..729M,2005Icar..173..362M} and a differential algebra based method \citep{2018MNRAS.479.5474L}, linear methods are still a matter of interest. The number of VAs' orbits, which have to be propagated in the Monte-Carlo method, is inversely proportional to the impact probability value making this method inefficient for not high impact probabilities. Line of variation method is more efficient, however it still needs sever thousands of orbits to be propagated. 
 {If the time of a collision is not far from the epoch of observations and} a two-body formalism can be applied using a linear method would be more rational. Also LOV method uses a linear method (namely the target plane method) so improvement of linear methods might lead to improvement of LOV method. 
 {It is worth noting that the application of the method is not limited to the asteroid and comet hazard problem. With little changes it can be used for many other similar problems, for instance for the problem of a satellite avoiding space debris.}

The paper is organized as follows. Section~\ref{sec:linear_methods} describes the principles  of linear methods for impact probability calculation. In particular,  the target plane method is presented in section~\ref{sec:target_plane} and section~\ref{sec:curvilinear_method} contains a brief description of the curvilinear coordinate system and the method which was fully described by \citet{2015MNRAS.446..705V}. Section~\ref{sec:novel_approach} presents a modification of this method making it more accurate, faster and easier to use. In section~\ref{sec:results} there is a practical comparison of three linear methods and discussion. This is followed by a conclusion in Section~\ref{sec:conclusion}.

\section{Linear methods}
\label{sec:linear_methods}

Impact probability is a consequence of our knowledge of the asteroid's orbit. If the orbit is precise the probability can be either zero or unity depending on whether the asteroid on a collisional course or not. However, in general, the asteroid orbit errors are not negligible, while the  Earth's orbit can be considered as precise in the considered problem \citep{2015jsrs.conf...92P}.

The errors of the asteroid's orbital parameters, in general, are assumed to have a normal distribution at the epoch of observations (epoch of orbital improvement). With time the orbital parameters' errors evolve and the uncertainty region grows. In linear methods it is also assumed that there is a linear relation between parameters' errors at the epoch of observations and at all considered epochs, making orbital parameters errors distribution to be normal at all considered epochs. This is a crucial assumption that is fulfilled only if the influence of gravitational perturbations from massive bodies (except the Sun) are almost the same for each part of the uncertainty region. In general, this means that either the uncertainty region is very small or the massive bodies' gravitation (except the Sun) is negligible. It should be noted that the chose of the orbital parameters is extremely important and different sets of orbital parameters gain different results.

\subsection{Target plane method}
\label{sec:target_plane}

Let us first define what a target plane is. 
The target plane (or b-plane) is a plane perpendicular to the incoming asymptote of the osculating geocentric hyperbola, or, equivalently, it is oriented normal to the unperturbed geocentric velocity $\bmath{v}_{\infty}$.  {It was introduced in astrodynamics by} \citet{1961P&SS....7..125K}  {and then by} \citet{1988Icar...75....1G}  {in the framework of \"{O}pik's theory} \citep{1976iecg.book.....O}.

In the target plane method for impact probability computation Cartesian coordinates and velocities are chosen as orbital parameters. Hence, the uncertainty region is represented as a 6-dimensional ellipsoid in coordinates and velocities. At the time $t$ when the asteroid comes close to the Earth (in general at a distance of the Earth's radius of   {action}) we can determine the uncertainty region by computing the covariance matrix. Let $\bmath{w}_0 = (x_0,y_0,z_0,\dot{x_0},\dot{y_0},\dot{z_0})$ be a vector of coordinates and velocities at the epoch of observations $t_0$ and $\mathbf{C}_0$ be the covariance matrix. Then the covariance matrix at time $t$ is:

\begin{equation}
\mathbf{C}_t = \mathbf{\Phi}(t_0,t) \ \mathbf{C}_0 \ \mathbf{\Phi^\mathrm{T}}(t_0,t), 
\end{equation}

\noindent where $\mathrm{T}$ denotes the matrix transpose operation and $\mathbf{\Phi}(t_0,t)$ is a matrix of partial derivations:

\begin{equation}
\mathbf{\Phi}(t_0,t) = \left(
\begin{array}{ccc}
\frac{\partial x }{\partial x_0} & \cdots &  \frac{\partial x }{\partial \dot{z}_0}\\
\vdots & \ddots & \vdots \\
\frac{\partial \dot{z} }{\partial x_0} & \cdots & \frac{\partial \dot{z} }{\partial \dot{z}_0} \\
\end{array}
\right) ,
\end{equation}

\noindent where $\bmath{w} = (x,y,z,\dot{x},\dot{y},\dot{z})$ is a vector of coordinates and velocities at time $t$. In general this matrix is computed while integrating the equations of motion of the asteroid with the variational equations \citep{1964Battin_book} or by a differential algebra techniques \citep{2014AdSpR..53..490M}.


\begin{figure}
	\includegraphics[width=\columnwidth]{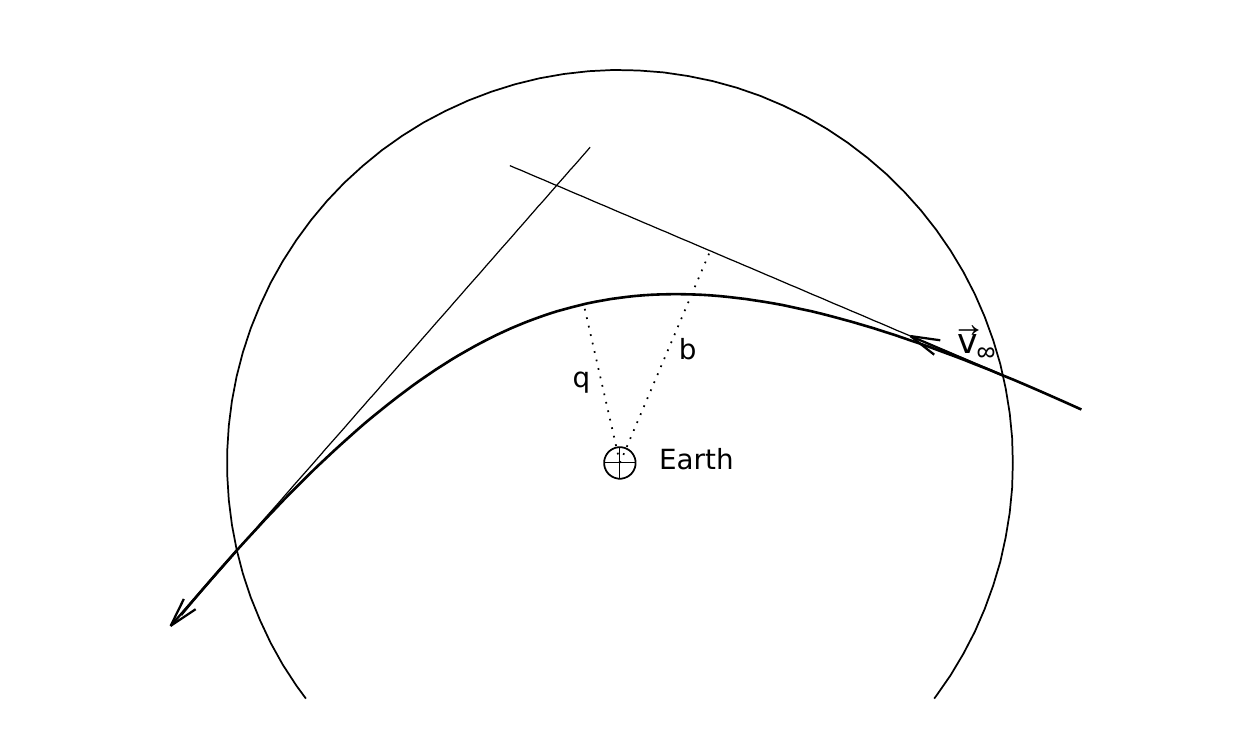}
	\caption{Schematically represented the asteroid's trajectory in the Earth's sphere of   {action}. $\bmath{v}_{\infty}$ is the asteroid's relative velocity when it enters the sphere of   {action}, $q$ is the minimum geocentric distance and $b$ is the so-called impact parameter, the distance from the geocenter and the hyperbola's asymptote.}
	\label{fig:Target_plane}
\end{figure}

The trajectory of an asteroid in the sphere of Earth's   {action} can be approximated by a hyperbola, as is shown in Fig.~\ref{fig:Target_plane}. The distance between the Earth, which is placed in the focus of the hyperbola, and the incoming asymptote is, in general, denoted $b$ and called impact parameter. Sometimes the target plane is called a 'b-plane' after this impact parameter. The minimum geocentric distance of the asteroid in this encounter depends on $b$ as:

\begin{equation}
q = \frac{b}{\sqrt{1+\frac{v_s^2}{v_{\infty}^2}}} ,
\end{equation}

\noindent where $\bmath{v}_s$ is the escape velocity from the Earth surface ($\approx 11.186$~km/s) and $\bmath{v}_{\infty}$ as before the unperturbed geocentric velocity. If the geocentric distance $q$ is less than the radius of the Earth then the asteroid will collide with the Earth. Therefore the collision happens if:

\begin{equation}
b \leq R_{\oplus} \sqrt{1+\frac{v_s^2}{v_{\infty}^2}},
\label{eq:b_param}
\end{equation}

\noindent where $R_{\oplus}$ is the radius of the Earth (here we approximate the Earth's shape as a sphere). A little before entering the sphere of   {action} the geocentric trajectories of the asteroid and its uncertainty region can be approximated by a straight line. The VAs, the impact parameters of which, satisfy inequation~(\ref{eq:b_param}) will collide with the Earth. To compute the impact probability we can project the uncertainty region on to the b-plane (which is perpendicular to the geocentric velocity of the region) and then compute the probability of impact as the probability of being closer to the project centre of the Earth than the right-hand part of inequation~(\ref{eq:b_param}). 

The projection on to the target plane can be represented by multiplication on a $2\times6$ matrix, $\mathbf{B}$. The coordinates on the target plane $\bmath{u} = \mathbf{B}\bmath{w}$. However, the 3 last columns of matrix $\mathbf{B}$ equal zero. The covariance matrix on the b-plane is given by: $\mathbf{L} = \mathbf{B} \mathbf{C} \mathbf{B}^\mathrm{T}$ \citep{1952Raobook}. Let the centre of coordinate system on the target plane be in the projection of the nominal asteroid. Then the impact probability is calculated as an integral of the probability-density function of the errors of $\bmath{u}$ over the projection of the Earth (a circle) with the radius $R'_{\oplus} = R_{\oplus} \sqrt{1+\frac{v_s^2}{v_{\infty}^2}}$:

\begin{equation}
P = \frac{1}{2\pi |\mathrm{det}\mathbf{L}|^{\frac{1}{2}}} \int \limits_{S_{R'_{\oplus}}} e^{-(\bmath{u}^\mathrm{T} \mathbf{L}^{-1} \bmath{u})/2} \mathrm{d} \bmath{u}.
\end{equation}

As it was mention above this method has several limitations. Not only the assumption of the linear relation between orbital parameters' errors at different epochs can be not fulfilled, but the nominal asteroid's orbit must also pass close to the Earth. Also if the uncertainty region is highly stretched, for instance, up to half of the orbit's length (the mean anomaly error is about 90$^\circ$) the distribution of coordinates and velocities errors can not be approximated by normal law. The possible collision can happen at one of the ends of the region and the target plane method would not work and even spot this possible collision.

\subsection{Method which uses a curvilinear system}
\label{sec:curvilinear_method} 

\begin{figure}
	\includegraphics[width=\columnwidth]{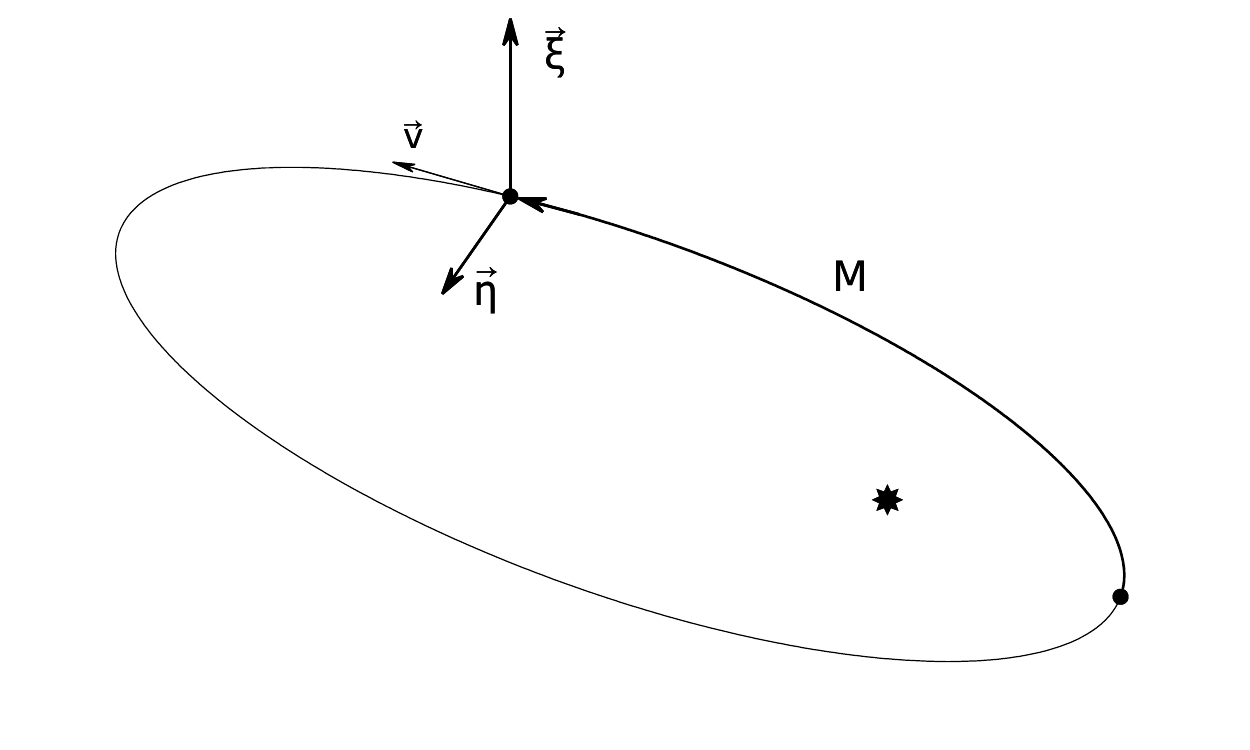}
	\caption{The curvilinear coordinate system related to the nominal orbit of an asteroid. The mean anomaly $M$ is an angle coordinate of the system, $\xi$ and $\eta$ are linear coordinates, the origin of which is the point corresponding to $M$.}
	\label{fig:Curv_coord_system}
\end{figure}

In order to take into account the fact that the uncertainty region is curved and stretched mostly along the nominal orbit \citet{2015MNRAS.446..705V} invented a new curvilinear coordinate system $ (\xi, \eta, M) $. Here we are presenting a general idea of this system. The system is constructed as follows. First of all, we fix the osculating orbit of a small body at time $t$ (i.e. the five parameters of the osculating ellipse). The mean anomaly $M$ in the osculating orbit is one of the coordinates of this system. The origin of the linear coordinates $\xi, \eta$ is the point on the ellipse corresponding to $M$. The $\xi$-axis is perpendicular to the plane of the fixed ellipse. The axis $\eta$ lies in the plane of the fixed ellipse and completes $ (\xi, \eta, M) $ to orthogonal, see Fig.~\ref{fig:Curv_coord_system}. 

A normal law  of errors of coordinates and velocities of this system  approximates better the actual distribution of VAs than a normal law in a Cartesian system. To compute the impact probability at time $t$ we have to find a covariance matrix in this system 

\begin{equation}
\mathbf{C}_{\xi \eta M \dot{\xi}\dot{\eta}\dot{M}} = \mathbf{Q} \cdot \mathbf{C}_{xyz\dot{x}\dot{y}\dot{z}} \cdot \mathbf{Q}^\mathrm{T},
\label{eq:Covar_xi} 
\end{equation}

\noindent where $\mathbf{C}_{xyz\dot{x}\dot{y}\dot{z}}$ is a covariance matrix in a Cartesian coordinate system at time $t$, and $\mathbf{Q}$ is a transfer matrix:

\begin{equation}
\mathbf{Q} = \left(
\begin{array}{ccc}
\frac{\partial \xi }{\partial x} & \cdots &  \frac{\partial \xi }{\partial \dot{z}}\\
\vdots & \ddots & \vdots \\
\frac{\partial \dot{M} }{\partial x} & \cdots & \frac{\partial \dot{M} }{\partial \dot{z}} \\
\end{array}
\right) .
\end{equation}

\citet{2015MNRAS.446..705V} calculated the impact probability at exact fixed time $t$ as a probability that the asteroid is closer than $R'_{\oplus}$ to the Earth. This can be written as:

\begin{equation}
\label{eq:6integral}
\frac{1} {\sqrt{(2\pi)^3} |\mathrm{det}\mathbf{C}_{\xi \eta M}|^{\frac{1}{2}}}  \int \limits_{\Theta} e^{-\frac{1}{2} (\bmath{w}^\mathrm{T} \mathbf{C}^{-1}_{\xi \eta M}\bmath{{w}})} \mathrm{d}\bmath{w},
\end{equation}

\noindent where $\Theta$ is a volume of the Earth in the curvilinear coordinate system, $\mathbf{C}_{\xi \eta M}$ is a $3 \times 3$ covariance matrix for  $(\xi, \eta, M)$ and $\bmath{w}$ is a three-dimensional vector of deviation from the nominal values of  $(\xi, \eta, M)$. Note that the nominal values are the values of the nominal asteroid, hence $\xi = \eta = \dot{\xi} = \dot{\eta} = 0$ for it. 

Using the curvilinear coordinate system for impact probability estimation problem was shown to be more robust in comparison to a Cartesian coordinate system.
However, one can see some of the disadvantages of the written above approach. First of all, the computed probability is the probability of a collision at exact time $t$. Since the time of a possible collision is unknown a set of probabilities are calculated with a small time step in the vicinity of the epoch when the Earth is at the minimum distance from the asteroid's orbit. It requires hundreds of integral~(\ref{eq:6integral}) computation. Because of that for the sake of efficiency \citet{2015MNRAS.446..705V} replaced the volume of the Earth $\Theta$ by a cuboid (rectangular parallelepiped) in system  $(\xi, \eta, M)$. This let the three-dimensional integral to be divided into a multiplication of three one-dimensional integrals making the method more efficient, but on the other hand, making the resulted impact probability higher. From the set of computed impact probabilities the one with the highest value was chosen. The chosen impact probability is not a cumulative one and can be less. Thus, the estimated this way impact probability value can differ by several times.

\section{A novel approach. Partial banana mapping}
\label{sec:novel_approach}

Here in this paper we are improving the method, which uses the curvilinear coordinate system, so that it would be more efficient,  will compute a cumulative probability and the volume of the Earth would not be replaced by a cuboid. The main disadvantage of the target plane method is approximation of the uncertainty region by an ellipsoid in a Cartesian coordinate system, which is not taking into account the curvature nature of the region. On the other hand the method, which uses the curvilinear coordinate system, does take into account the fact that the uncertainty region looks like a 'bananoid', but because of the absence of projection in the method the resulted impact probability is not cumulative. Here we are combining these two approaches. However, it should be noted that instead of projecting the whole 'bananoid' we need to project only the part of it which is the closest to the Earth. That is why the new approach is suggested to be called 'partial banana mapping'.

\begin{figure}
	\includegraphics[width=\columnwidth]{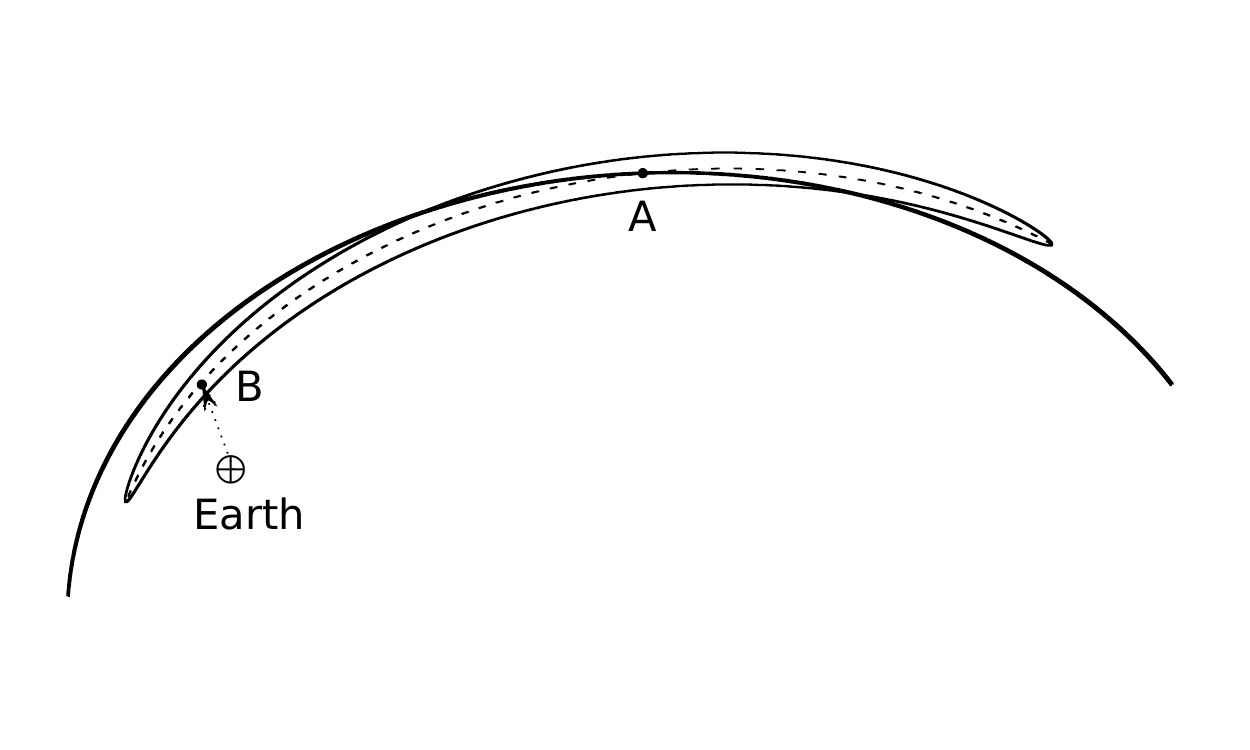}
	\caption{The schematic illustration of the confidence curvilinear ellipsoid. Point 'A' is the nominal position of the asteroid, point 'B' is the VA on the main axis of the confidence ellipsoid, which is closest to the Earth after projection on to its target plane. The arrow indicates the Earth's velocity direction with respect to the confidence ellipsoid. The bold line is the nominal asteroid's orbit.}
	\label{fig:Closest_aster}
\end{figure}

At the approximate time of a possible collision we calculate the covariance matrix $\mathbf{C}_{\xi \eta M \dot{\xi}\dot{\eta}\dot{M}}$ from equation~(\ref{eq:Covar_xi}). In general this time is close to the time when the Earth is at the closet point to the nominal orbit of the asteroid. However, it does not coincide exactly, since the uncertainty region is not exactly elongated along the nominal orbit (see Fig.~\ref{fig:Closest_aster}). Then on the main axis of the curvilinear confidence ellipsoid we find that VA, which is going to be the closest to the Earth after projection on to the target plane. It should be noted that each VA has its own target plane and we consider the six-dimensional confidence ellipsoid including velocities. 

\subsection{Finding the closest VA}
\label{subsec:findingVA}
The covariance matrix $\mathbf{C}_{\xi \eta M \dot{\xi}\dot{\eta}\dot{M}}$ is a positive definite matrix. Hence, we can use a spectral decomposition of this matrix:

\begin{equation}
\mathbf{C}_{\xi \eta M \dot{\xi}\dot{\eta}\dot{M}} = \mathbf{V} \cdot \mathbf{\Lambda} \cdot \mathbf{V}^\mathrm{T}, 
\end{equation}

\noindent where $\mathbf{V}$ is an orthogonal matrix composed of eigen vectors of matrix $\mathbf{C}_{\xi \eta M \dot{\xi}\dot{\eta}\dot{M}}$ and $\mathbf{\Lambda}$ is a diagonal matrix with eigen values on a diagonal. 




Let the eigen value in the first row and first column be the maximal one. Let $\bmath{w}_0 = (\xi_0, \eta_0, M_0, \dot{\xi}_0, \dot{\eta}_0, \dot{M}_0)$ be the curvilinear coordinates of the nominal asteroid ($\xi_0 = \eta_0 = \dot{\xi}_0 = \dot{\eta}_0 = 0$) and $\bmath{w}$ is a six-dimensional random vector in the curvilinear coordinate system. The covariance matrix of $\bmath{w}$ is $\mathbf{C}_{\xi \eta M \dot{\xi}\dot{\eta}\dot{M}}$. $\mathbf{\Lambda}$ is a covariance matrix for the random vector $\bmath{u} = \mathbf{V}^\mathrm{T} (\bmath{w} - \bmath{w}_0)$  \citep{1952Raobook}. Since $\mathbf{\Lambda}$ is a diagonal matrix if the last five components of vector $\bmath{u}$ equal zero ($\bmath{u} = (u_1,0,0,0,0,0)$) it defines a VA on the main axis of the curvilinear six-dimensional confidence ellipsoid.

With a given  $u_1$ one can find coordinates and velocities of the corresponding VA and compute the distance between this VA and the Earth's centre after projection on to its target plane. Hence the distance from the Earth's centre on a target plane to VAs from the main axis of the curvilinear confidence ellipsoid is a one variable function (function of $u_1$). Using, for instance, the golden-section search technique \citep{1953Kiefer} one can find $u_1^*$ which corresponds to VA with the minimal distance.

\subsection{Impact probability estimation}

Let us notice that in a small region (several radii of the Earth) the curvature of the uncertainty region can not be recognized. Consequently in the small vicinity of the found closest VA the uncertainty region can be well approximated by a part of an ellipsoid in a Cartesian coordinate system. To construct this region properly we do the following. Let as above $u_1^*$ be the first coordinate of vector $\bmath{u}$ corresponding to the found closest VA on the main axis of the curvilinear confidence ellipsoid. Let  $(x_*,y_*,z_*,\dot{x}_*,\dot{y}_*,\dot{z}_*)$ and $(\xi_*, \eta_*, M_*, \dot{\xi}_*, \dot{\eta}_*, \dot{M}_*)$ be the coordinates and velocities of the found VA in a Cartesian and the curvilinear coordinate systems correspondingly. If the found VA is not close to entering the Earth's sphere of   {action} we should propagate the orbit of the nominal asteroid to the time when the closest VA enters the sphere of   {action}. The transfer matrix is:

\begin{equation}
\mathbf{Q}_* = \left(
\begin{array}{ccc}
\frac{\partial \xi_* }{\partial x_*} & \cdots &  \frac{\partial \xi_* }{\partial \dot{z}_*}\\
\vdots & \ddots & \vdots \\
\frac{\partial \dot{M}_* }{\partial x_*} & \cdots & \frac{\partial \dot{M}_* }{\partial \dot{z}_*} \\
\end{array}
\right) .
\end{equation}

The covariance matrix of Cartesian coordinates and velocities corresponding to the found closest VA can be found by:

\begin{equation}
\mathbf{C}_{xyz\dot{x}\dot{y}\dot{z}}^* = \mathbf{Q}_*^{-1} \cdot \mathbf{C}_{\xi \eta M \dot{\xi}\dot{\eta}\dot{M}} \cdot {\mathbf{Q}_*^\mathrm{T}}^{-1}.
\end{equation}

Then we compute an impact probability $P_*$ for the found VA using its Cartesian coordinates and velocities  $(x_*,y_*,z_*,\dot{x}_*,\dot{y}_*,\dot{z}_*)$ and covariance matrix $\mathbf{C}_{xyz\dot{x}\dot{y}\dot{z}}^*$ by a target plane method. The obtained impact probability then should be corrected since $P_*$ computed not for the nominal asteroid but for a virtual asteroid which can be significantly far along the main axis of the curvilinear confidence  ellipsoid. The final impact probability is:

\begin{equation}
P = \mathrm{e}^{-\frac{1}{2} ( u_1^* / \sqrt{\lambda_{11}} )^2} P_*
\label{eq:correction}
\end{equation}

\noindent where $\lambda_{11}$ is an element of matrix $\mathbf{\Lambda}$ in the first row and first column, which is a dispersion for the first component of vector $\bmath{u}$.

\section{Results}
\label{sec:results}

\begin{table}
	\caption{Characteristics of asteroids' orbits}
	\label{table:asteroids_parameters}
	\begin{center}
		\begin{tabular}{l|c|c|c|r|r}
			Designation &   impact time       & $\sigma,^{\prime \prime}$    & $T_{1}$      & $\Delta T$, d &  $k$ \\
			\hline                                                                  
			2006 JY26  &    2073-05-03.4      & 0.69                         & 2006-05-06.4      & 3.81          & 77 \\
			2006 QV89  &    2019-09-10.5      & 0.44                         & 2006-08-29.3      & 10.83         & 68 \\
			2010 UK    &    2068-09-15.4      & 0.57                         & 2010-10-17.5      & 14.71         & 81 \\	
			2011 AG5   &    2040-02-05.2      & 0.36                         & 2010-11-08.6      & 316.77        & 213 \\
			2007 VK184 &    2048-06-03.1      & 0.39                         & 2007-11-12.1      & 60.01         & 102\\
			2007 VE191 &    2015-11-27.1      & 0.45                         & 2007-11-15.3      & 13.25         & 68 \\
			2008 JL3   &    2027-05-01.4      & 0.40                         & 2008-05-05.4      & 4.10          & 31 \\
			2014 WA    &    2049-11-16.1      & 0.82                         & 2014-11-16.3      & 0.60          & 53 \\
			2009 JF1   &    2022-05-06.3      & 0.55                         & 2009-05-04.4      & 1.23          & 25 \\
			2012 MF7   &    2046-06-21.9      & 0.40                         & 2012-06-23.4      & 29.06         & 27 \\
			2008 CK70  &    2030-02-14.7      & 0.44                         & 2008-02-09.3      & 4.66          & 77 \\
			2005 BS1   &    2016-01-14.4      & 0.57                         & 2005-01-16.3      & 3.13          & 25 \\
			2005 QK76  &    2030-02-26.3      & 0.65                         & 2005-08-30.2      & 1.68          & 14 \\
			2007 KO4   &    2015-11-23.2      & 0.49                         & 2007-05-22.3      & 3.00          & 14 \\
		\end{tabular}
	\end{center}
	
	\medskip
	'Designation' is the asteroid's designation, 'impact time' is the date of impact that was computed from the partial banana mapping method, 
	$\sigma$ is the root mean square of observations, $T_1$ is the time of the first observations,
	$k$ is the number of observations, $\Delta T$ is the observation arc.
	
\end{table}

\begin{table*}
	\caption{Results}
	\label{table:comparison_unlinear}
	\begin{center}
		\begin{tabular}{l|c|c|c|c|c|c|c|c|c}
			Designation  &  $P_{MC} \pm 3\sigma_{MC}$    &   $P_{TP}$           &  $P_{\xi \eta M}$    & $P_{PBM}$           & $P_{PBM_{\boxplus}}$ &$\Delta M, ^\circ$ &$\Delta T_{TP}, d$&$\Delta T_{PBM}, d$ &     COP            \\
			\hline                                                                                                                                         
			2006 JY26    & $(5.6\pm 1.7)\cdot 10^{-5} $  & $ 1.0\cdot 10^{-4}$  &$1.1 \cdot 10^{-4} $  & $ 1.0\cdot 10^{-4}$ &  $ 1.1\cdot 10^{-4}$  &$0.77$    & $2.2$        & $12.4$  & $ 3.2\cdot 10^{-2}$\\
			2006 QV89    & $(1.8\pm 0.1)\cdot 10^{-3} $  & $ 2.0\cdot 10^{-3}$  &$2.2 \cdot 10^{-3} $  & $ 1.8\cdot 10^{-3}$ &  $ 2.6\cdot 10^{-3}$  &$0.85$    & $0.4$        & $7.4$   & $ 5.6\cdot 10^{-3}$\\
			2010 UK      & $(3.1\pm 0.7)\cdot 10^{-3} $  & $ 1.8\cdot 10^{-3}$  &$2.7 \cdot 10^{-3} $  & $ 1.8\cdot 10^{-3}$ &  $ 2.7\cdot 10^{-3}$  &$1.22$    & $0.4$        & $4.6$   & $ 5.4\cdot 10^{-3}$\\
			2011 AG5     & $(5.3\pm 1.3)\cdot 10^{-4} $  & $ 4.2\cdot 10^{-4}$  &$5.1 \cdot 10^{-4} $  & $ 5.7\cdot 10^{-4}$ &  $ 6.0\cdot 10^{-4}$  &$0.90$    & $0.4$        & $9.0$   & $ 3.9\cdot 10^{-3}$\\
			2007 VK184   & $(6.2\pm 2.0)\cdot 10^{-6} $  & $ 2.7\cdot 10^{-5}$  &$3.0 \cdot 10^{-5} $  & $ 2.6\cdot 10^{-5}$ &  $ 3.5\cdot 10^{-5}$  &$0.96$    & $0.4$        & $6.6$   & $ 3.4\cdot 10^{-3}$\\
			2007 VE191   & $(6.4\pm 1.0)\cdot 10^{-4} $  & $ 0.0             $  &$6.3 \cdot 10^{-4} $  & $ 6.8\cdot 10^{-4}$ &  $ 7.7\cdot 10^{-4}$  &$0.66$    & $0.0$        & $4.8$   & $ 1.1\cdot 10^{-3}$\\
			2008 JL3     & $(3.0\pm 0.4)\cdot 10^{-4} $  & $ 7.5\cdot 10^{-4}$  &$4.7 \cdot 10^{-4} $  & $ 4.0\cdot 10^{-4}$ &  $ 1.0\cdot 10^{-3}$  &$0.17$    & $0.2$        & $3.0$   & $ 1.0\cdot 10^{-3}$\\
			2014 WA      & $(3.5\pm 2.4)\cdot 10^{-7} $  & $ 0.0             $  &$4.5 \cdot 10^{-7} $  & $ 5.4\cdot 10^{-7}$ &  $ 7.0\cdot 10^{-7}$  &$5.02$    & $0.0$        & $3.0$   & $ 9.8\cdot 10^{-4}$\\
			2009 JF1     & $(7.4\pm 1.2)\cdot 10^{-4} $  & $ 7.3\cdot 10^{-4}$  &$6.6 \cdot 10^{-4} $  & $ 8.0\cdot 10^{-4}$ &  $ 8.1\cdot 10^{-4}$  &$0.14$    & $3.2$        & $6.8$   & $ 8.2\cdot 10^{-4}$\\
			2012 MF7     & $(3.1\pm 0.8)\cdot 10^{-4} $  & $ 0.0             $  &$4.0 \cdot 10^{-4} $  & $ 4.8\cdot 10^{-4}$ &  $ 5.1\cdot 10^{-4}$  &$2.37$    & $0.0$        & $2.8$   & $ 8.2\cdot 10^{-4}$\\
			2008 CK70    & $(6.4\pm 1.0)\cdot 10^{-4} $  & $ 5.8\cdot 10^{-4}$  &$6.4 \cdot 10^{-4} $  & $ 6.9\cdot 10^{-4}$ &  $ 6.9\cdot 10^{-4}$  &$0.85$    & $1.0$        & $6.6$   & $ 7.2\cdot 10^{-4}$\\
			2005 BS1     & $(1.4\pm 0.2)\cdot 10^{-4} $  & $ 0.0             $  &$1.5 \cdot 10^{-4} $  & $ 1.4\cdot 10^{-4}$ &  $ 2.0\cdot 10^{-4}$  &$1.44$    & $0.0$        & $5.4$   & $ 2.1\cdot 10^{-4}$\\
			2005 QK76    & $(4.3\pm 0.9)\cdot 10^{-5} $  & $ 0.0             $  &$3.8 \cdot 10^{-5} $  & $ 4.1\cdot 10^{-5}$ &  $ 4.5\cdot 10^{-5}$  &$10.6$    & $0.0$        & $4.6$   & $ 4.6\cdot 10^{-5}$\\
			2007 KO4     & $(7.3\pm 4.0)\cdot 10^{-7} $  & $ 0.0             $  &$4.0 \cdot 10^{-7} $  & $ 2.1\cdot 10^{-6}$ &  $ 2.8\cdot 10^{-6}$  &$27.6$    & $0.0$        & $7.2$   & $ 1.6\cdot 10^{-5}$\\
		\end{tabular}
	\end{center}
	
	\medskip
	\textit{Notes:} 'Designation' is the asteroid's designation, $P_{MC}$ is the probability computed by the Monte-Carlo method, $\sigma_{MC}$ is a standard deviation of $P_{MC}$, $P_{TP}$, $P_{\xi \eta M}$  and $P_{PBM}$ are probabilities calculated by the target plane method, method, which uses a curvilinear coordinate system, and the partial banana mapping method correspondingly.  $P_{PBM_{\boxplus}}$ is the probability computed by the partial banana mapping method, but the Earth's projection is replaced by a square. $\Delta M$ is a difference in the mean anomaly between the nominal asteroid and the found virtual asteroid which is the closest to the Earth. COP is the probability that the asteroid would have if it was on a collisional trajectory but with the same covariance matrix.
\end{table*}

To verify the new approach the impact probabilities were computed for a set of asteroids by different methods.  {The results obtained by the Monte-Carlo method are considered as etalon values.} The asteroids were chosen randomly from the website of the Jet Propulsion Laboratory, NASA\footnote{\url{https://cneos.jpl.nasa.gov/sentry/.}}, but preferably with not low impact probabilities. This set of asteroids includes asteroids from the previous paper \citep{2015MNRAS.446..705V} in order to use the values obtained by the Monte-Carlo method. Also, some new asteroids were added.   {Some information about these asteroids is presented in Table~\ref{table:asteroids_parameters}.} It should be noted that nowadays for some of the asteroids from the list new observations are available. Using new observations a new asteroid's orbit and a covariance matrix can be computed that can drastically change the impact probability value. Thus it can be said that this comparison is made on model examples which are, however, based on real ones.

All the results are presented in Table \ref{table:comparison_unlinear}. The standard deviation of the Monte-Carlo method's results are calculated according to $\sqrt{P_{MC}(1-P_{MC})}/\sqrt{m}$, where $m$ is the number of realizations, $P_{MC}$ is the Monte-Carlo's probability. In the table along with the probabilities computed by the partial banana mapping method ($P_{PBM}$), the target plane method ($P_{TP}$) and method, which uses $(\xi, \eta, M)$ coordinate system ($P_{\xi \eta M}$),  we give the probabilities computed by the partial banana mapping method but with the Earth's projection replaced by a square ($P_{PBM_{\boxplus}}$). This helps us understand the reason for some inconsistencies between $P_{\xi \eta M}$ and $P_{PBM}$. $\Delta M$ is the difference in the mean anomaly between the nominal asteroid and the VA, which was found in section~\ref{subsec:findingVA}.
The impact probability value can slightly depend on the epoch at which the covariance matrix is computed.  {(For the target plane method the time of covariance matrix computation generally is the time when the asteroid enters the Earth's sphere of  action).} To show this dependence we also present in the table the time interval in which the probability changes less than 10 per cent of its value for the target plane method ($\Delta T_{TP}$) and the partial banana mapping method ($\Delta T_{PBM}$). 

The last column is the Collisional Orbit Probability value (COP), which is the probability that the asteroid would have if it was on a collisional trajectory. It is also the maximum value of impact probability the asteroid can get with this covariance matrix. This value is a great indicator of the size of the six-dimensional confidence ellipsoid. A similar value has already been used in the work of \citet{2006IAUS..229..215C} where it was called 'maximum possible impact probability'. It should be also noted that the value of COP  slightly depends on the method which used to calculate it. For instance, the target plane method and the partial banana mapping method use different virtual asteroids, the velocities of which are used to projection. Hence, the difference in the velocities' directions leads to the difference in the COP value,  {however, the difference should not be very high}. Here we present the value of COP obtained by the partial banana mapping method by simply taking $u_1^* = 0$ in formula (\ref{eq:correction}) and the asteroid's coordinates after projection the same as the Earth's coordinates. The asteroids in Table~\ref{table:comparison_unlinear} are in descending order of COP.

From Table~\ref{table:comparison_unlinear} one can see that $\sigma_{MC}$ values are small which means the probabilities $P_{MC}$ given by the Monte-Carlo method are accurate enough to be considered as etalon values. The classical target plane method didn't find possible collisions for six of the considered asteroids. Also the nominal asteroid came to the sphere of  {action} only in the cases of asteroids 2008~JL3 and 2009~JF1.  {Besides there is no case where the classical target plane method shows more accurate values of impact probability than the partial banana mapping method}. These results justify our assumptions. The lesser the value of COP (the higher the uncertainty region) and the higher $\Delta M$  the higher the likelihood that the target plane method would not work properly. The values of $P_{\xi \eta M}$ and $P_{PBM}$ are close to each other in general, however, sometimes they can differ by several times as we expected. If the asteroid's coordinates on the target plane are not close to the Earth's ones, replacing the Earth's projection by a square can significantly change the impact probability. In these cases $P_{\xi \eta M}$ can be higher than $P_{PBM}$, for instance the cases of 2006~QV89, 2010~UK and 2008~JL3. The difference between $P_{PBM}$ and $P_{PBM_{\boxplus}}$ shows how significant the replacing can be. On the other hand, if the coordinates are close to each other but the angle between the relative velocity's direction and the main axis of the confidence ellipsoid is small then because of the projection in the partial banana mapping method, $P_{PBM}$ can be higher than $P_{\xi \eta M}$. For instance the cases of 2014~WA, 2009~JF1 and 2007~KO4. 

The 8-th and 9-th columns show how accurate we should be in choosing the epoch, at which the covariance matrix is calculated, for the target plane and partial banana mapping methods correspondingly. As one can see $\Delta T_{PBM}$ take values starting from several days, which means the epoch of the covariance matrix computation can be defined with the accuracy of a couple of days. In contrast this epoch for the target plane method must be much more accurate. 

 {From Table~\ref{table:comparison_unlinear} it is seen that $P_{PBM}$ values are in good agreement with the results of the Monte-Carlo method and do not differ from them by more than several times.} However, the partial banana mapping is still a linear method, which means it has its limitations of usage. As it was mentioned above to use this method there should be a linear relation between errors of $(\xi,\eta,M,\dot{\xi},\dot{\eta},\dot{M})$ at considered epochs. This assumption is well fulfilled if a two-body formalism can be applied or the influence of gravitational perturbations from massive bodies (except the Sun) should be almost the same for each part of the uncertainty region. 

 {This assumption can be drastically violated if the asteroid comes close to a massive body (e.g. a major planet). This happened for some of the asteroids from our list. Asteroid 2006 JY26 had several close approaches with the Earth from the epoch of observations till the possible collision time. The closest ones were on 2 November 2009 and 4 November 2010 when the asteroid passed the Earth at distances of 0.017~au and 0.037~au respectively. Asteroid 2010~UK also had several close approaches with massive planets. On 17 June 2031 it came closer than 0.05~au to Venus and on 25 October 2035 it came up to 0.03~au to the Earth. Asteroid 2007~VK184 approached Venus on 1 June 2030 at a distance of 0.065~au but also it approached the Earth at a distance of 0.037~au three days before the possible collision. All these close approaches affected the resulted impact probability values. It should be noted that for these three asteroids the results of the partial banana mapping method coincides with the results of the classical target plane method, which justifies that the discrepancy in the impact probability values are because of the close approaches.}

 {The situation with asteroid 2007~KO4 is more interesting. It had only one close approach with the Earth on 20 November 2007 with a distance of 0.10~au, which is further than for the asteroids mentioned above. Probably the lesser the impact probability value the stronger close approaches affect on the result. This is indirectly confirmed by comparing the results of CLOMON~2}\footnote{\url{https://newton.spacedys.com/neodys/index.php?pc=4.1.}} and Sentry\footnote[1]{\url{https://cneos.jpl.nasa.gov/sentry/.}}  {impact monitoring systems. The lesser the impact probability value the more the chance that the results are different. Higher impact probabilities are more resistant to small changes. However, the influence of exact values of close approach distance on different values of impact probabilities computed by linear methods are beyond this paper and is a problem for future investigations.}
But if the assumption of linear relation is fulfilled the partial banana mapping is an efficient method for impact probability computation since it requires propagation of only one asteroid orbit and is more robust than the target plane method.

\section{Conclusions}
\label{sec:conclusion}

The paper presents a new linear method for impact probability estimation of asteroids with the Earth. This method is a significant improvement of the method, which uses a special curvilinear coordinate system. This system is associated with the asteroid's nominal orbit. One of the coordinates of the system is the mean anomaly on the osculating orbit. The other two coordinates are projections on to axes $\xi$ and $\eta$, the origin of which is in the point on the osculating orbit corresponding to given mean anomaly. The axes are perpendicular to each other and to the osculating orbit. Each virtual asteroid can be represented as the coordinates and velocities of this system. Assumption of a normal distribution of  errors of the curvilinear coordinates and velocities of an asteroid describes well  the actual distribution of virtual asteroids, which looks like a region that curved and stretched mainly along the nominal asteroid's orbit.

Since the distribution of virtual asteroids is mainly stretched along the nominal asteroid's orbit, a collision can happen when the Earth is close to the asteroid's orbit (note that the Earth can be extremely far from the nominal asteroid at that time). In the method, which uses the curvilinear system, the impact probability is computed as a three dimensional integral at a set of times close to the time when the Earth is closest to the asteroid's orbit. For the sake of computational efficiency the volume of the Earth is replaced by a cuboid which let the three-dimensional integral to be decomposed to a multiplication of three one-dimensional integrals. The highest probability is chosen to be the solution. In this case the given probability is the probability of a collision at exact time and not the cumulative probability. 

In the new method, which is proposed to be called 'partial banana mapping method', the probability computation also starts when the Earth is at the closest position to the asteroid's orbit. Among the virtual asteroids on the main axis of the curvilinear confidence ellipsoid the one, which is closest to the Earth after projection on to its target plane, is found. If the found virtual asteroid is not close to entering the Earth's sphere of   {action} the orbit of the nominal asteroid is propagated till it does. The part of the curvilinear confidence ellipsoid, around the found virtual asteroid, is obtained and projected on to its target plane. Then the impact probability of the virtual asteroid with the Earth is calculated by the classical target plane method. Then this probability is multiplied by a factor of $\mathrm{e}^{-\sigma^2 /2}$, which is the value of the probability density function for the found virtual asteroid. This approach has several significant advantages. The probability is computed for only one time and not for a set of times and it is cumulative in the vicinity of the Earth's approach to the asteroid's orbit. Also the volume of the Earth is not replaced by a cuboid. Comparison with the classical target plane method made on 14 examples showed the advantage of the new method.  {The results are in good agreement with the Monte-Carlo method's results.} However, it should be mentioned that this is still a linear method and the assumption of linear relation of errors of coordinates and velocities is important. If the gravitation from other massive bodies (except the Sun) is significant and different parts of the confidence region are disturbed differently none of the linear methods would work.

\section*{Acknowledgements}

The author is grateful to Dr. Siegfried Eggl from University of Washington for his help with inventing the name for the 'partial banana mapping' method. This work was supported by the Russian Scientific Foundation grant No. 16-12-00071.




\bibliographystyle{mnras}
\bibliography{PBM_ARXIV} 








\bsp	
\label{lastpage}
\end{document}